%Paper: hep-ph/9204228
%From: Peter Arnold <arnold@galileo.phys.washington.edu>
%Date: Wed, 22 Apr 92 17:17:51 PDT

%
% ***NOTE***
%
% It's a sad, unfortunate world and so, if you actually want to see the
% figures, you'll have to write and request a snail-mail copy from me.
%
%                                 Peter Arnold
%                                 arnold@washington.phys.galileo.edu
%

\input harvmac

%%%%%%%%%%%%%%%%%%%%%% temporary, until harvmac fixed:

\def\refmark#1{${}^{\refs{#1}}$\ }
\def\footsym{*}\def\footsymbol{}\ftno=-2
\def\foot{\ifnum\ftno<\pageno\xdef\footsymbol{}\advance\ftno by1\relax
\ifnum\ftno=\pageno\if*\footsym\def\footsym{$^\dagger$}\else\def\footsym{*}\fi
\else\def\footsym{*}\fi\global\ftno=\pageno\fi
\xdef\footsymbol{\footsym\footsymbol}\footnote{\footsymbol}}
%%%%%%%%%%%%%%%%%%%%%%%%%%%%%%%%%%%%%%%%%%%%%%%%%%%%%%%%%%%%%%%%%%%%%%%%%

\font\authorfont=cmcsc10 \ifx\answ\bigans\else scaled\magstep1\fi

\baselineskip=0.29in plus 2pt minus 2pt

\def\d{{\rm d}}
\def\Tc{T_{\rm c}}
\def\ML{M_{\rm L}}
\def\MT{M_{\rm T}}
\def\mh{m_{\scriptscriptstyle\rm H}}
\def\mt{m_{\rm t}}
\def\Veff{V_{\rm eff}}
\def\Vring{V_{\rm ring}}
\def\absPhi{\left|\Phi\right|}
\def\gy{g_{\rm y}}
\def\mueff{\mu_{\rm eff}}
\def\meff{m_{\rm eff}}
\def\Rxi{R$_\xi$}
\def\phivev{\langle\phi\rangle}
\def\phicl{\phi^{\rm cl}}
\def\lsim{ \vcenter{\hbox{$\buildrel{\displaystyle <}\over\sim$}} }
\def\gsim{ \vcenter{\hbox{$\buildrel{\displaystyle >}\over\sim$}} }

\lref\lindea{D.A. Kirzhnits and A.D. Linde, \sl Phys.\ Lett.\ \bf 42B\rm,
  471 (1972).}
\lref\weinberg{S. Weinberg, \sl Phys.\ Rev.\ \bf D9\rm, 2257 (1974).}
\lref\dolan{L. Dolan and R. Jackiw, \sl Phys.\ Rev.\ \bf D9\rm, 3320 (1974).}
\lref\dolanb{L. Dolan and R. Jackiw, \sl Phys.\ Rev.\ \bf D9\rm, 1686 (1974).}
\lref\linde{D.A. Kirzhnits and A.D. Linde, \sl Ann.\ Phys. \bf 101\rm,
  195 (1976).}
\lref\takahashi{K. Takahashi, \sl Z. Phys.\ \bf C26\rm, 601 (1985).  The
  reader should beware that the gauge theory cases are handled incorrectly
  in this paper.}
\lref\fendley{P. Fendley, \sl Phys.\ Lett.\ \bf B196\rm, 175 (1987).}
\lref\carrington{M.E. Carrington, University of Minnesota preprint
  TPI-Minn-91/48-T (December 1991).  The reader should beware that her
  eq.\ (37) is meant to read ${5\over3}g'^2T^2$ for $\pi_\psi^{(1)}(0)$
  and $g^2T^2$ for $\pi_\psi^{(2)}(0)$.}
\lref\parwani{An example of consistently using ring-improvements to
  higher orders in perturbation theory may be found in
  R.R. Parwani, SUNY preprint ITB-SB-91-64 (December 1991).
  This paper examines the effective scalar mass (and not the general
  effective potential) in an unbroken scalar theory.}
\lref\brahm{D.E. Brahm and S.D.H. Hsu, Cal. Tech. preprint CALT-68-1705
  (December 1991).}
\lref\shaposhnikov{M.E. Shaposhnikov, CERN preprint CERN-TH.6319/91
  (November 1991).}
\lref\ginsparg{P. Ginsparg, \sl Nucl.\ Phys.\ \bf B170\rm, 388 (1980).}
\lref\arnold{P. Arnold, E. Braaten and S. Vokos, Argonne National Lab
  preprint ANL-HEP-PR-91-108 (December 1991).}
\lref\gpy{For a review, see
  D.J. Gross, R.D. Pisarski and L.G. Yaffe, \sl Rev.\ Mod.\ Phys.\ \bf 53\rm,
  43 (1981).}

\Title{
  \vbox{\baselineskip12pt
    \hbox{UW/PT-92-06}
    \hbox{NUHEP-TH-92-06}}
  }{
  \vbox{
    \centerline{Phase Transition Temperatures}
    \centerline{at Next-to-Leading Order}
  }}
\centerline{\authorfont Peter Arnold}
\centerline{\sl Department of Physics, University of Washington,
    Seattle, WA  98195\foot{permanent address}}
\centerline{\sl Department of Physics and Astronomy, Northwestern University,
    Evanston, IL 60208}

\vskip .3in
Broken gauge symmetries are typically restored at high temperature,
and the leading-order result for the critical temperature $\Tc$ was
found many years ago by Weinberg and by Dolan and Jackiw.  I find
a simple expression for the next-to-leading order correction to $\Tc$,
which is order $e\Tc$ where $e$ is the gauge coupling.
The result is a simple consequence of
recent work on summing ring diagrams at high temperature in gauge
theories.  The result is valid when the Higgs self-coupling $\lambda$
is the same order as $e^2$, and it does not
address the case of strongly first-order phase transitions, which arise
when $\lambda \ll e^2$.

%\draft
\Date{March 1992}

%\draftmode

\sequentialequations

In simple Higgs models of spontaneously broken gauge theories, the classical
potential of the Higgs field has the usual form
\eqn\Vcl{
   V(\phi) = -{1\over2}\mu^2\phi^2 + {1\over4!}\lambda\phi^4 .
}
At high temperatures, however, there is an additional effective
mass term\refmark{\lindea\weinberg\dolan-\linde}
of order $e^2 T^2\phi^2$ which is analogous to the Debye screening mass in an
electromagnetic plasma.  The effective potential at high temperature is
given approximately by
\eqn\Vleading{
   \Veff(\phi) \approx {1\over2} (-\mu^2 + c e^2 T^2) \phi^2
      + {1\over4!}\lambda\phi^4 ,
}
where the constant $c$ is model dependent.

In any given model, the
critical temperature for symmetry restoration is then determined to
leading order by the effective potential \Vleading:
\eqn\Tcleading{
    \Tc \approx {\mu\over\sqrt{c} e} .
}
I shall show that recent improvements to the effective potential yield
simple corrections of order $e\Tc$ to this formula for $\Tc$.  Explicit
formulas will be presented for the Abelian Higgs model and for
the weak sector of the minimal standard model.

The procedure is quite simple.  In the next section, I review the
ring-improved effective
potential\refmark{\dolan\linde\takahashi\fendley\carrington-\parwani},
which has recently been implemented by Carrington\refmark\carrington\
for gauge theories such as the
Standard Model.
In section 3, I then find the critical temperature
by requiring that $\d^2\Veff/\d^2\phi = 0$ at $\phi = 0$.  Finally,
I argue that corrections to this result are higher-order than $e\Tc$.
In particular,
finding $\Tc$ by requiring the curvature of $\Veff$ to vanish at the
origin is appropriate for a second-order phase transition but not
for a first-order one; I show that the resulting discrepancy is higher order
than $e\Tc$ provided $\lambda \sim e^2$.
Working in general Lorentz gauges, I check that
the result is independent of the gauge parameter.
I also find that the naive ring-improved potential would generate further
corrections to $\Tc$ at order $e^{3/2}\Tc$.  These corrections are a
manifestation of the failure of the naive ring approximation at this order.
A simple improvement of the approximation reduces the corrections
to order $e^2\Tc$, at which point there is no known perturbative method
for calculating them.
In an Appendix, I discuss the subtleties
of finding the same results in \Rxi\ gauges.

\newsec{Review of the Ring-Improved Effective Potential}

\subsec{Pure scalar theory}

For simplicity, start by ignoring the gauge fields and focus on the simple
theory of a single, real scalar field with the symmetry breaking potential
\Vcl.
Diagrammatically, the screening mass discussed above
comes from the quadratically divergent loop shown in
\fig\fquadscalar{
  The order $\lambda T^2$ contribution to the squared effective mass
  in a simple scalar theory.
}.
After subtraction of zero-temperature counter-terms, the quadratic divergence
is cut off by T, so that such diagrams are order $\lambda T^2$.
At high temperatures, this is the dominant interaction of the
full one-loop potential.  It has the effect of replacing $-\mu^2$ in
the classical potential by
\eqn\eqmueff{
  -\mueff^2 = -\mu^2 + {1\over24}\lambda T^2 ,
}
giving $\Tc^2 \approx 24\mu^2/\lambda$.

The full one-loop potential consists of all interactions generated at
one-loop, such as
\fig\fVoneloop{
  A generic contribution to the full one-loop effective potential.
}.
For general theories, each species of particle gives
a contribution to the finite-temperature piece of the one-loop potential
that is simply the free energy of an ideal gas of such particles.
Restricting attention to bosons:\refmark{\dolan}
\eqn\Voneloopa{
    \Veff(T,\phi) = \Veff(0,\phi) + \sum_i n_i \Delta V_i(T,\phi) ,
}
\eqn\Voneloopb{
   \Delta V_i(T,\phi) =
   T\int\nolimits {\d^3k\over(2\pi)^3}
   \ln\left\{1 - \exp[-\beta\sqrt{k^2+m_i^2(\phi)}]\right\} ,
}
where $\beta$ is the inverse temperature, the sum is over all species $i$,
and $n_i$ is the number of degrees of freedom associated with each species.
$m_i(\phi)$ is the
effective mass of species $i$ in the presence of a background scalar
field $\phi$.
The high temperature limit $(T \gg m_i(\phi))$ yields:
\eqn\DVexpand{
      \Delta V(T,\phi) = {\rm const.} ~+~ {1\over24} m^2(\phi) T^2
         - {1\over12\pi} m^3(\phi) T ~+~ O(m^4\ln T) .
      \qquad (\rm bosons)
}
The constant above is temperature dependent but $\phi$ independent;
it is not relevant to determining the critical temperature
and shall be ignored.

For the simple scalar theory, there
is only one species --- the Higgs --- and $m^2(\phi)$ is the second
derivative of the classical potential:
\eqn\mHiggs{
   m^2_{\rm cl}(\phi) = -\mu^2 + {1\over2} \lambda \phi^2 .
}
However, we have already seen that the effective value of $\mu^2$ at
finite temperature is quite different from the classical value
at high temperature ($T\,\gsim\,\Tc$).
When studying temperatures of order $\Tc$,
it is therefore important\refmark{\dolan\linde\takahashi\fendley-\carrington}
to make the replacement \eqmueff\ and use
instead
\eqn\mHiggs{
   \meff^2(\phi) = -\mu^2 + {1\over24}\lambda T^2 + {1\over2}\lambda\phi^2
}
in the one-loop potential \DVexpand.  This substitution corresponds to
including the dominant contributions of the ring (also known as daisy)
diagrams shown in
\fig\Vdaisy{
  A generic contribution to the ring-improved one-loop effective potential.
  The small loops are hard with loop momenta $\sim T$; the large loop is
  soft, except for the special case of \fquadscalar.
}.

The leading (non-constant) term in the expansion
\DVexpand\ simply reproduces the
dominant $\lambda T^2 \phi^2$ interaction discussed earlier.
The next term, of order $m^3(\phi) T$, is the term that will generate the
first correction to $\Tc$ in gauge theories.  For the pure scalar theory,
the ring-improved potential in this expansion is
\eqn\scalarring{
  \eqalign{
    \Vring(\phi) =
     {1\over2} \left(-\mu^2 + {1\over24}\lambda T^2\right) \phi^2
  &
     - {T\over12\pi}
       \left(-\mu^2 + {1\over24}\lambda T^2 + {1\over2}\lambda\phi^2
       \right)^{3/2}
  \cr &
     + {1\over4!} \lambda \phi^4
     + O(m^4 \ln T) .
  }
}

It will later be convenient to also view the ring-improved potential in the
language of decoupling and the renormalization group.
At high temperature, loops which are less than quadratically divergent
(or non-divergent pieces of quadratically divergent ones) are dominated by
their infrared behavior.
In Euclidean space, this means loop momenta
are dominated\refmark\weinberg\ by
$k_0=0$ and $|\vec k| \sim m$.  The dominant $k_0=0$ piece of the
finite-temperature frequency sum $T\sum_{k_0}$ gives such loops a linear,
rather than quadratic, dependence on $T$.  Taking $k_0=0$ in all such loops
yields an effective three-dimensional
theory whose squared coupling is $\lambda T$ and which may be viewed
as an approximate effective theory at scales much smaller than $T$.
The contributions from physics at scale $T$ will decouple like powers
of $1/T$ except for possible renormalizations of masses and so forth.
The replacement \eqmueff\
is a statement of the relation between the renormalized mass
$-\mueff^2$ in the effective three-dimensional theory and the
renormalized mass $-\mu^2$ in the zero-temperature theory.  Now computing
and renormalizing the simple one-loop potential in the effective
three-dimensional theory gives
\eqn\Vthree{
  \eqalign{
    \Delta V(T,\phi) &= {T\over12\pi}
       (-\mueff^2 + {1\over2}\lambda\phi^2)^{3/2} ,
  \cr
    \Veff(T,\phi) &\approx -{1\over2}\mueff^2\phi^2 + {1\over4!}\lambda\phi^4
                    + \Delta V(T,\phi)
  \cr}
}
which is equivalent, within my approximations, to the ring-improved result
\scalarring.

What is the size of corrections to the ring-improved
one-loop potential from other diagrams?  The squared coupling in the
three-dimensional theory is $\lambda T$, so each loop added costs a factor
of $\lambda T/\meff$.
The effective value of $m$ approaches zero as $T$ approaches
$\Tc$, and so the ring-improved loop expansion will break down very close to
the phase transition, when $|\meff|\,\lsim\,\lambda T$.
Eq.~\eqmueff\ implies this
breakdown occurs when $|T-\Tc|\,\lsim\,\lambda\Tc$, and so there is no
simple way to compute corrections of order $\lambda\Tc$ to $\Tc$.

\subsec{Abelian Higgs Model}

Now focus on the simplest example of a spontaneously broken
gauge theory: the Abelian Higgs model given by
\eqn\lzero{
  {\cal L} = - {1\over4}F^2 + \left|D\Phi\right|^2 - V(\absPhi^2) ,
}
\eqn\vZero{
  V(\absPhi^2) = - \mu^2\absPhi^2 + {1\over3!}\lambda\absPhi^4 ,
}
where $\Phi$ is a complex field and
$D_\mu\Phi = (\partial_\mu - ieA_\mu)\Phi$.
I shall typically express the potential in terms of
$\phi = \absPhi/\sqrt2$, so that it takes the form \Vcl,
and shall work in Lorentz gauges, where the gauge fixing
term is
\eqn\lgf{
  {\cal L}_{\rm g.f.} = {1\over\xi} (\partial\cdot A)^2 .
}
I shall assume $\lambda \sim e^2$ unless stated otherwise.

Now consider the one-loop effective potential \Voneloopa.  In Landau
gauge ($\xi=0$), the mass squared in the presence of a background scalar field
$\phi$ is classically
\eqn\masses
{
  \matrix{
     M^2(\phi) = e^2\phi^2 \hfill
          & \quad ({\rm vector}) \hfill \cr
     \vphantom{\biggl[}
     m_1^2(\phi) = -\mu^2+{1\over2}\lambda\phi^2 \hfill
          & \quad ({\rm physical~Higgs}) \hfill \cr
     m_2^2(\phi) = -\mu^2+{1\over6}\lambda\phi^2 \hfill
          & \quad ({\rm unphysical~Goldstone~boson}) \hfill \cr
   }
}
The unimproved one-loop potential in
Landau gauge is then
\eqn\Vunimproved{
  \eqalign{
    \Veff(\phi) \approx
      {1\over2} & \left[ -\mu^2 + \left({\lambda\over2} + {\lambda\over 6}
         + 3e^2 \right) {T^2\over12} \right] \phi^2
   \cr &
      - {T\over12\pi} \left[ 3e^3\phi^3
         + (-\mu^2+{1\over2}\lambda\phi^2)^{3/2}
         + (-\mu^2+{1\over6}\lambda\phi^2)^{3/2} \right]
   \cr &
      + {1\over4!} \lambda\phi^4 .
  }
}
\nfig\fquadvector{
  The order $e^2 T^2$ contribution to the squared effective scalar mass.
}
To make the ring improvement, we need the leading finite-temperature
contributions to the effective particle masses.  For the Higgs boson,
it can be read from the first term of \Vunimproved\ and corresponds to
figs.~\xfig\fquadscalar\ and \xfig\fquadvector.
It is the same
for the unphysical Goldstone boson:\foot{
  Technically, these substitutions only make sense inside IR dominated
  loops, where the loop momentum $k$ is $\ll T$.  The self-energies
  $\Pi(k)$ coming from the hard thermal loops of \fquadvector\
  can then be
  approximated by $\Pi(0)$ when constructing ring-improved propagators.
  The substitutions are {\it not} valid
  in the $m^2(\phi)T^2$ term of \DVexpand\ which, unlike the
  subsequent terms, arises from the quadratic divergence of loops, where
  the loop momentum $k$ is order $T$.
  In the case at hand, such worries only affect the
  $\phi$-independent constant terms, which I am ignoring.
  The linear terms in the effective potential found in
  Refs.~\refs\brahm\ and \refs\shaposhnikov,
  however, are the result of higher-order versions of such substitutions
  improperly made into the $m^2(\phi)T^2$ term.
}
\eqn\mtscalar{
  \eqalign{
     &m_1^2(\phi) \rightarrow -\mu^2
            + \left({2\over3}\lambda + 3e^2\right){T^2\over12}
            + {1\over2}\lambda\phi^2 \cr
     &m_2^2(\phi) \rightarrow -\mu^2
            + \left({2\over3}\lambda + 3e^2\right){T^2\over12}
            + {1\over6}\lambda\phi^2 \cr
  }
}
The leading contribution to the thermal vector mass comes from the
diagrams of
\fig\fvmass{
  Order $e^2 T^2$ contributions to the squared thermal vector mass.
}
and is momentum dependent.  However ring graphs
of the form of \fig\fVvdaisy{
   A generic daisy graph contribution to the effective potential in
   the Abelian Higgs model.  The main loop is dominated by its infrared
   behavior.}
are dominated by their Euclidean infrared
behavior,\foot{
  The exception is the ultraviolet piece of the quadratic interactions
  like figs.~\xfig\fquadscalar\ and \xfig\fquadvector\ with {\it no} mass
  insertions.  These pieces give
  the $e^2 T^2 \phi^2$ interactions and are independent of the particle
  masses.  The pieces of these and other diagrams which {\it do} depend
  on the particle masses, however, are dominated by their infrared behavior.}
corresponding to momenta $k_0 = 0$ and $|\vec k| \ll T$.
In this limit, the diagrams of \fvmass\ generate a Debye
screening mass of $eT/\sqrt{3}$ for $A_0$ (the longitudinal polarization)
and nothing at the same order for $\vec A$.  So, for computing the
ring-improvement to the effective potential:
\eqn\mtvector{
  \matrix{
     \ML^2(\phi) \rightarrow e^2\phi^2 + {1\over3}e^2T^2 \hfill
            & ({\rm longitudinal~polarization}), \hfill \cr
     \vphantom{\biggl[}
     \MT^2(\phi) \rightarrow e^2\phi^2 \hfill
            & ({\rm transverse~polarizations}), \hfill \cr
  }
}
\eqn\Vabelian{
  \eqalign{
    \Vring(\phi) \approx
      {1\over2} \meff^2(T) \phi^2
      - {T\over12\pi} \biggl[ &
         (e^2\phi^2 + {1\over3}e^2T^2)^{3/2}
         + 2 e^3\phi^3
         + (\meff^2(T) + {1\over2}\lambda\phi^2)^{3/2}
  \cr &
         + (\meff^2(T) + {1\over6}\lambda\phi^2)^{3/2} \biggr]
      + {1\over4!} \lambda\phi^4 ,
  }
}
where
\eqn\mbardef{
  \meff^2(T) = -\mu^2
     + \left( {2\over3}\lambda + 3e^2 \right) {T^2\over12} .
}

In general Lorentz gauge, one must include the unphysical polarization
(the polarization proportional to the four-momentum) of the photon.
When the background scalar field $\phi$ is non-zero, this polarization
mixes with the unphysical Goldstone boson.  Taking the one-loop potential
 from Ref.~\refs\dolan\ and incorporating the ring improvement gives
\eqn\Vlorentz{
  \eqalign{
    \Vring(\phi) \approx
      {1\over2} \meff^2(T) \phi^2
      - {T\over12\pi} \biggl[
  &
         (e^2\phi^2 + {1\over3}e^2T^2)^{3/2}
         + 2 e^3\phi^3
         + (\meff^2(T) + {1\over2}\lambda\phi^2)^{3/2}
  \cr &
         + R_+^3
         + R_-^3 \biggr]
      + {1\over4!} \lambda\phi^4 ,
  \cr}
}
where
\eqn\Rpm{
  R_\pm^2 = {1\over2} \bar m_2^2
    \pm {1\over2} \sqrt{ \bar m_2^2 (\bar m_2^2 - 4\xi e^2\phi^2) },
}
\eqn\mtilde{
  \bar m_2^2 = \meff^2(T) + {1\over6} \lambda \phi^2 .
}

\newsec{The Critical Temperature}

With the ring-improved potential in hand, consider the computation of
the critical temperature.  The curvature of the effective potential
\Vlorentz\ at $\phi=0$ is\foot{
  Instead of expanding around arbitrary background $\phi$ to compute
  $\Veff(\phi)$, I could have obtained $\Veff''(0)$ by simply expanding about
  $\phi=0$ as usual and then evaluating the two-point function by
  ring-improving diagrams such as
  figs.~\xfig\fquadscalar\ and \xfig\fquadvector.
  Having $\Veff(\phi)$,
  however, is useful for the later discussion of first vs.\ second order
  phase transitions and for the study of \Rxi\ gauge in the appendix.}
\eqn\dVtwo{
  \Vring''(0) = \meff^2(T) - {\sqrt{3}\over12\pi} e^3T^2
      - \left({\lambda\over6\pi} - {\xi e^2\over4\pi}\right)
            \meff(T) T .
}
Solving $\Vring''(0)=0$ to order $e\Tc$, I find that the last term
in \dVtwo\ is irrelevant.  The result is
\eqn\abelianTc{
  \Tc^2 = {\mu^2 \over {1\over12}\left({2\over3}\lambda+3e^2\right)
          - {\sqrt{3}\over12\pi} e^3 }
       ~+~ O(e^{3/2}\Tc) }
and is independent of the gauge parameter $\xi$ to this order.
The source of the order $e\Tc$ correction to the leading-order result
is the photon Debye screening mass, which generated the second term
in $\dVtwo$.

\subsec{Validity of expansion}

 From the review of pure scalar theory, we know that the loop expansion
breaks down when $|T-\Tc| \,\lsim\, e^2\Tc$, and so the critical
temperature cannot be easily computed within $e^2\Tc$.  To determine if
the value \abelianTc\ of $\Tc$ is correct to order $e\Tc$, one needs to
know if the formula \dVtwo\ is a good approximation when $|T-\Tc| \sim e\Tc$.
For such temperatures, $\meff^2 \sim e^3 T^2$.
Unfortunately, the argument is slightly complicated because the
naive ring-improved loop expansion is not controlled simply by $e^2T/m$,
as it was in the pure scalar case, because there are {\it two} soft
scales when $\phi$ is near zero: $m$ and $\ML\sim eT \gg m$.
(In order to avoid
more proliferation of scales in the following discussion, I shall
continue to focus on the effective potential close to $\phi = 0$.)

Look at the corrections to the result for $\Tc$ derived
using the ring-improved one-loop potential.  First note that $\Vring$
itself
implied a gauge-dependent correction of order $e^{3/2}\Tc$, which arises
 from the last term of \dVtwo.  The cause of this correction is that
$\meff^2(T)$ as defined by \mbardef\ is order $e^3\Tc^2$ (instead of
zero) at $\Tc$, and so the last term in \dVtwo\ is suppressed by only
$\sqrt{e}$ relative to the second term.
The correction arises because the $\meff^2(T)$ used in ring-improved
Higgs propagators is not small enough when
$|T-\Tc| \,\lsim\, e\Tc$---it is a poor approximation to $\Veff''(0)$.
One may fix the approximation by self-consistently
replacing $\meff^2(T)$ in the improved one-loop potential by the
leading terms of $V''(0)$ from \dVtwo:
\eqn\meffnew{
  \meff^2(T) \rightarrow -\mu^2
         + \left({2\over3}\lambda+3e^2\right) {T^2\over12}
         - {\sqrt{3}\over12\pi} e^3T^2 .}
Eq.~\dVtwo\ for $V''(0)$
then no longer produces an $O(e^{3/2}\Tc)$ contribution to $\Tc$.
Diagrammatically, the redefinition of $\meff(T)$
corresponds to using the dominant pieces
of once-iterated daisy graphs such as
\fig\fVdoubledaisy{
  A generic once-iterated daisy graph for the Higgs loop contribution to
  the effective potential.  The medium-size loops are longitudinal photons.
}
for the Higgs-loop
contribution to the potential.
The smallest loops in \fVdoubledaisy\ are hard, with momenta of order $T$.
The next-smallest loops are soft longitudinal photon loops, with momenta of
order $\ML\sim eT$.  Close to the critical temperature,
the large Higgs loop is softer yet, with momenta
of order $\meff \ll e\Tc$.  Because of the hierarchy of scales, it is
a good approximation at each level of \fVdoubledaisy\ to approximate
resummed propagators $1/[p^2+\Pi(p)]$ by $1/[p^2+\Pi(0)]$.

For the purpose of understanding the size of other corrections to the
effective potential, it is useful to restate the redefinition of
$\meff(T)$ in the language of decoupling.  I earlier reviewed the
effective three dimensional theory one obtains at scales below $T$.
But when the Higgs mass gets sufficiently light as we approach $\Tc$,
there is then another heavy scale in the problem---the longitudinal
photon mass of order $eT$---and a new effective theory may be obtained
by integrating out its effects.  The effects of this heavy scale will
by suppressed except for renormalization of masses and so forth.
The significant mass renormalization comes from the loop of
\fig\fLdecouple{
  The dominant scalar mass term induced by heavy ($eT$) longitudinal
  photons.
},
which gives the $e^3T^2$ term incorporated into $\meff^2$ in
\meffnew.
Effective interactions generated by the {\it heavy} contributions
(all loop momenta of order $eT$)
of higher-loop graphs will be suppressed by
$e^2T/\ML \sim e$ and affect the derivation of
$\Tc$ only at order $e^2\Tc$.

In the new effective theory, soft loops will be suppressed by
$e^2T/m$ and the loop expansion is controlled when $m \gg e^2 T$.
As an example, the two-loop graph shown in
\fig\fVtwoloop{
  An example of a two-loop contribution to the scalar mass.
}
is order
$e^4 T^2$, which affects $\Tc$ only at order $e^2\Tc$.

In conclusion, the error in the formula for \abelianTc\ is order $e^2\Tc$
rather than $e^{3/2}\Tc$.

\subsec{First vs. second-order transitions}

The one-loop ring approximation \Vlorentz\ to the effective potential
actually describes a first-order rather than second-order phase
transition.  As depicted in
\fig\forder{
  The ring-improved effective potential shown qualitatively at (1) the
  critical temperature $\Tc$ and (2) the temperature $T_0$ where
  $\Veff''(0) = 0$.  The curves have been arbitrarily normalized so
  that $\Veff(0) = 0$.
},
the critical temperature $\Tc$
for a first-order phase transition is different from the temperature
$T_0$ at which $V''(0)=0$.  The difference between these two temperatures
is easily estimated.  Working near the critical temperature, ignore
the $R_\pm^3$ terms in the potential and consider the form of the potential
for small values of $\phi$.  Then
\eqn\Vlimit{
  \Vring(\phi) \rightarrow
     {1\over2} \left( \meff^2(T) - {\sqrt{3}\over12\pi} e^3 T^2 \right) \phi^2
     - {1\over6\pi} e^3 T \phi^3 + {1\over4!}\lambda\phi^4 ,
}
where I have assumed that $\lambda\gg e^4$ so that corrections to
$\lambda\phi^4$ may be ignored.
At the true $\Tc$ depicted in \forder, all three terms above must be
the same order of magnitude.  Equating the magnitude of the last two terms
gives $\phi_c \sim e^3 T/\lambda$ and then equating with the first term
gives
\eqn\Tdiff{
  V''(0) \sim V''(\phi_c) \sim {e^6/\lambda} T^2,
  \qquad
  \Tc - T_0 \sim {e^4\over\lambda} \Tc .
}

For $\lambda \sim e^2$, the difference between the two temperatures
is order $e^2 \Tc$ and so does
not affect the earlier result for the order $e\Tc$ correction to $\Tc$.
Note that $\Tc$ is close enough to $T_0$ that $\meff^2 \sim e^2T$ at
$T_c$, and so the improved loop expansion has just
started to break down.  Thus, one may not conclude based solely on
the ring-improved effective potential that the phase transition is in
fact first order when $\lambda \sim e^2$.  The potential merely indicates
that, if it {\it is} first-order, then the difference between
$\Tc$ and $T_0$ is smaller than order $e\Tc$.  Other arguments, given by
Ginsparg,\refmark\ginsparg\
suggest that the transition is indeed first order.

For $\lambda \ll e^2$, the first-order nature of the phase transition
becomes strong enough that the effective potential can be trusted to
distinguish between first and second-order transitions.
In particular,
$\lambda \sim e^3$ implies $\Tc - T_0 \sim e\Tc$, and the
effective scalar mass is order $e^{3/2}T$ at $\Tc$, giving a loop
expansion controlled by $e^2T/m \sim \sqrt{e}$.  My earlier calculation
still gives $T_0$ to order $e\Tc$, but now this is not an accurate
calculation of $\Tc$ to the same order.  In this case, the formula
for $\Tc$ to order $e\Tc$ is not simple, and $\Tc$ is most easily found
by evaluating the ring-improved potential numerically, as was done by
Carrington.\refmark\carrington

\newsec{The Minimal Standard Model}

The calculation of the previous section is easily generalized to the
weak sector of the
Minimal Standard Model with three families, where
Carrington\refmark\carrington\
has derived
the ring-improved one-loop potential in Landau gauge.
Expanding her result in the high-temperature limit gives
\eqn\Vweak{
  \eqalign{
    \Vring(\phi) \approx
      {1\over2}m^2(T) \phi^2
   &
      - {T\over12\pi} \biggl[
          3\left({1\over4}g^2\phi^2 + {11\over6}g^2T^2\right)^{3/2}
          + 6 \left({1\over4}g^2\phi^2\right)^{3/2}
   \cr&\qquad
          + \left({1\over4}g'^2\phi^2+{11\over6}g'^2T^2\right)^{3/2}
          + 2 \left({1\over4}g'^2\phi^2\right)^{3/2}
   \cr&\qquad
          +   \left(m^2(T)+{1\over2}\lambda\phi^2\right)^{3/2}
          + 3 \left(m^2(T)+{1\over6}\lambda\phi^2\right)^{3/2}
       \biggr]
   \cr&
       + {1\over4!}\lambda\phi^4 ,
   \cr}
}
where
\eqn\mweak{
   \bar m^2(T) = -\mu^2
     + \left(\lambda+{9\over4}g^2+{3\over4}g'^2+3\gy^2\right){T^2\over12} .
}
$\gy$ is the top quark coupling, which is the only one I have treated
as significant.
The conventions for the coupling constants are that
$D_\mu = \partial_\mu + g A_\mu \cdot \tau/2 + Yg'B_mu/2$ for doublets,
the hypercharge is normalized so that $Q = T_3 + Y/2$,
and the Yukawa coupling is
$\gy\,\bar q_{\rm\scriptscriptstyle L}\!\cdot\!\Phi
t_{\rm\scriptscriptstyle R}$
where $\Phi$ is the full complex doublet,
whose classical potential is of the form \vZero.
The correction to $\Tc$ is generated by the Debye screening masses of
the gauge bosons, and proceeding as before gives
\eqn\Tcweak{
  \Tc^2 = {\mu^2 \over
     {1\over12}\left(\lambda+{9\over4}g^2+{3\over4}g'^2+3\gy^2\right)
     - {1\over12\pi}\sqrt{11\over6}\left({9\over4}g^3+{3\over4}g'^3\right)
  } ~~+~~ O(g^2 \Tc) .
}

In non-Abelian theories, a little more needs to be said about the convergence
of the loop expansion than in the Abelian case.  Because of the 3-point
gauge coupling, it is possible to construct loops solely from the massless
(at $\phi=0$),
transverse gauge bosons, such as contained in
\fig\ftransverse{
  A two-loop contribution from transverse gauge bosons to the effective
  scalar mass.
}.
Such loops are
generally infrared divergent.
It is presumed\refmark\gpy\
that such loops are cut-off by a
non-perturbative magnetic screening mass of order $g^2 T$, for which the
loop expansion parameter $g^2 T/\MT$ is then order 1.  However, if we
indeed cut off the infrared behavior of transverse gauge loops at
order $g^2 T$,
then their contributions to $V''(0)$, such as in \ftransverse, are order
$g^4 T^2$.  So the incalculable contribution of such loops only affects
$\Tc$ at order $g^2\Tc$.

As an example of the numeric size of the order $g\Tc$ corrections,
consider the effect of including or eliminating the $g^3$ and $g'^3$
terms in the denominator of \Tcweak.  The effect of the cubic terms is
largest when the Higgs and top masses are small; in the limit that
these masses are negligible (in which case \Tcweak\ may no longer be
valid), the inclusion of the cubic terms increases the result
for $\Tc$ by 37\%.  For $\mh=\mt=100$ GeV, including the cubic terms
increases the result for $\Tc$ by 13\%.

\vskip 0.5in

This work was supported by an SSC Fellowship from the Texas National
Research Laboratory Commission and by the U.S. Department of
Energy, grant DE-FG06-91ER40614.
I am indebted to Eric Braaten, and especially to Larry Yaffe, for many
long and useful discussions.

\appendix{A}{Results in \Rxi\ gauges}

Discussions of the effective potential in the literature sometimes employ
a generalized version of \Rxi\ gauge.  In this appendix, I discuss
how the previous analysis of $\Tc$ works in these gauges.  For simplicity,
I shall work in the Abelian Higgs model.

The generalized $R_\xi$ gauge of Dolan and Jackiw\refmark{\dolan,\dolanb}
is fixed
by
\eqn\fix{
  {\cal L}_{\rm g.f.} = - {1\over2\xi} (\partial_\mu A^\mu + \xi e v \phi_2)^2
    - \bar\eta (\partial^2 + \xi e^2 v \phi_1) \eta ,
}
where the complex Higgs field is decomposed as
$\Phi = (\phi_1 + i\phi_2)/\sqrt2$, $\eta$ is the ghost field,
and $v$ is an additional, arbitrary gauge parameter.
In the usual definition of $R_\xi$ gauge,
$v$ is set to the classical value $\phicl$ about which one expands
$\phi_1$; this definition eliminates mixing of the scalars with
the gauge field.  This definition is unacceptable for computing the
effective potential $\Veff(\phicl)$, however, because the effective potential
is gauge dependent.
With the usual definition of $v$, varying $\phicl$ corresponds to
changing gauge and so $\Veff$ will not be consistently
computed in a single gauge.  This pathology destroys the equality between the
derivatives of $\Veff$ and zero-momentum tadpoles, self-energies
and so forth.
Following Dolan and Jackiw, I shall instead fix $v$ independent of
$\phicl$ at the cost of having mixing between $\phi$ and $A_\mu$.
Note that $v=0$ is the previously treated case of Lorentz gauges.

Rather than studying the full potential $V(\phicl_1,\phicl_2)$, I
shall simplify the discussion by restricting attention to
$\phicl_2 = 0$.  With this restriction, mixing occurs between the
unphysical polarization (proportional to $k_\mu$) of $A_\mu$ and
the unphysical Goldstone boson, but there is no mixing with the
physical Higgs.
I shall also generally drop subscripts and so forth
to write $V(\phicl_1,0)$ as $V(\phi)$.

As discussed in Ref.~\refs\arnold, it is important that gauge parameters not
take extreme values (such as $\xi\,\gsim\,1/e^2$); otherwise, the loop
expansion for the effective potential breaks down in a non-trivial
manner, making it difficult to compute $\Tc$ even to leading-order in
such gauges.  For this discussion, I shall assume $\xi \sim 1$.
Requiring that the loop expansion be well-behaved then also puts
constraints on how large $v$ can be.
In particular, consider the diagonal element, of the non-diagonal
propagator, that propagates the unphysical polarization of $A_\mu$ into
itself.  Expanding about $\phi=0$ and taking the limit
$v\rightarrow\infty$, this component of the propagator turns out to be
\eqn\gaugeprop{
  {p_\mu p_\nu \over p^2} {\xi^2 e^2 v^2 \over p^2 (p^2 + m^2)}
}
where $m$ is the effective Higgs mass.  Loops involving this propagator
will then be enhanced by factors of $\xi^2e^2v^2/m^2$ if $v$ is large.
To avoid enhancement of higher-loop graphs, which would invalidate
the improved one-loop approximation to the potential, I shall
restrict attention to gauges with $v\,\lsim\,m/e$.

The ring-improved one-loop potential is
\eqn\Vrxi{
  \eqalign{
    \Vring(\phi) \approx
      {1\over2} \meff^2(T) \phi^2
      - {T\over12\pi} \biggl[
  &
         (e^2\phi^2 + {1\over3}e^2T^2)^{3/2}
         + 2 e^3\phi^3
         + (\meff^2(T) + {1\over2}\lambda\phi^2)^{3/2}
  \cr &
         + R_+^3 T
         + R_-^3 T
         - 2(\xi e^2v\phi)^{3/2} \biggr]
      + {1\over4!} \lambda\phi^4 ,
  \cr}
}
where the last term inside the brackets is from the ghost contribution
and where
\eqn\Rpmxi{
  R_\pm^2 = {1\over2} (\bar m_2^2 + 2\xi e^2v\phi)
    \pm {1\over2} \sqrt{ \bar m_2^2 (\bar m_2^2 - 4\xi e^2(\phi-v)\phi) },
}
\eqn\mtildexi{
  \bar m_2^2 = \meff^2(T) + {1\over6} \lambda \phi^2 .
}
Taking $V''(0)$ in this approximation gives an infinite result.
This is not disastrous
because $\phi=0$ is not, in fact, the symmetric state in this
gauge.  Expanding $\Vring(\phi)$ about $\phi = 0$ gives
\eqn\Vexpand{
  \eqalign{
    \Vring(\phi) \rightarrow
      {\rm const.}
   &
      - \left( {1\over4\pi} \xi e^2 v T\meff(T) \right) \phi
   \cr&
      + {1\over2} \left[ \meff^2(T) - {\sqrt{3}\over12\pi} e^3T^2
         - \left({\lambda\over6\pi} - {\xi e^2\over4\pi}\right)
         \right] \phi^2
   \cr&
      + \cdots .
   \cr}
}
There is a linear term which above $\Tc$ shifts the ground state away from
$\phi=0$.  This is not surprising because the \Rxi\ gauge choice
\fix\ explicitly breaks the global isospin rotation and
$\phi\rightarrow -\phi$ symmetries which distinguish $\phi=0$
 from other values of $\phi$; the symmetric vacuum need not be at
$\phi = 0$.\foot{
  Above $\Tc$, there should be no magnetic screening in this model.
  The reader may wonder how a $\phi\not=0$ vacuum can give a zero photon
  mass.  The $e\phivev$ contribution to the photon mass
  is the same order as 2-loop corrections to that mass, against which it
  presumably cancels.  I have not checked this cancellation explicitly.}

If higher-order terms in \Vexpand\ are ignored, then $V''(\phi)$ in the
shifted vacuum is just the coefficient of the quadratic term in
\Vexpand.  The analysis is then the same as it was for the earlier case
of Lorentz gauges, and one finds the same result for $\Tc$ to order
$e\Tc$.  In this approximation, the ground state above $\Tc$ is at
$\phivev$ of order $e^2 v T/\meff$.
The correction from higher-order terms in \Vexpand\ is
suppressed by $e^2v\phivev/\meff^2$ which, using my restriction
that $v\,\lsim\,\meff/e$ is a suppression by at least the usual
expansion parameter $e^2 T/\meff$.

\listrefs
\listfigs

\end